\documentclass[aip,reprint]{revtex4-1}

\newcommand{\lb}{\label}
\newcommand{\be}{\begin{equation}}
\newcommand{\ee}{\end{equation}}
\newcommand{\ber}{\begin{eqnarray}}
\newcommand{\eer}{\end{eqnarray}}
\newcommand{\bers}{\begin{eqnarray*}}
\newcommand{\eers}{\end{eqnarray*}}

\usepackage{graphicx}
\usepackage{dcolumn}
\usepackage{bm}

\begin{document}





\title{Relation of Astrophysical Turbulence and Magnetic Reconnection} 








\author{A. Lazarian}
\affiliation{Department of Astronomy, University of Wisconsin, 475 North 
Charter Street, Madison, WI 53706, USA}

\author{Gregory L. Eyink}
\affiliation{Department of Applied Mathematics \& Statistics, The Johns Hopkins University 
University, Baltimore, MD 21218, USA}
\altaffiliation{also Department of Physics \& Astronomy}

\author{E. T. Vishniac}
\affiliation{Department of Physics and Astronomy, McMaster University, 1280 Main Street West, 
Hamilton, ON L8S 4M1, Canada}











\date{\today}

\begin{abstract}

Astrophysical fluids are generically turbulent and this must be taken into account for most transport processes. We discuss how the preexisting turbulence modifies magnetic reconnection and how magnetic reconnection affects the MHD turbulent cascade. We show the intrinsic interdependence and interrelation of magnetic turbulence and magnetic reconnection, in particular, that strong magnetic turbulence in 3D requires reconnection and 3D magnetic turbulence entails fast reconnection. We follow the approach in Eyink, Lazarian \& Vishniac 2011 to show that the expressions of fast magnetic reconnection in Lazarian \& Vishniac 1999 can be recovered if Richardson diffusion of turbulent flows is used instead of ordinary Ohmic diffusion. This does not revive, however, the concept of magnetic turbulent diffusion which assumes that magnetic fields can be mixed up in a passive way down to a very small dissipation scales. On the contrary, we are dealing the reconnection of dynamically important magnetic field bundles which strongly resist bending and have well defined mean direction weakly perturbed by turbulence. We argue that in the presence of turbulence the very concept of flux-freezing requires modification. The diffusion that arises from magnetic turbulence can be called reconnection diffusion as it based on reconnection of magnetic field lines. The reconnection diffusion has important implications for the continuous transport processes in magnetized plasmas and for star formation. In addition, fast magnetic reconnection in turbulent media induces the First order Fermi acceleration
of energetic particles, can explain solar flares and gamma ray bursts. However, the most dramatic
consequence of these developments is the fact that the standard flux freezing concept must be
radically modified in the presence of turbulence.
\end{abstract}


\maketitle 



\section{Purpose and Outline}

The purpose of this short paper is to discuss processes that govern the change of the magnetic field topology in astrophysical fluids. We claim that it is incorrect to ignore ubiquitous astrophysical turbulence while studying magnetized reconnection. We also show the intrinsic and very deep relation between magnetic reconnection and turbulence.

In what follows, we discuss the problem of astrophysical reconnection and point out the difference that exists in terms of reconnection between astrophysical systems and their present day numerical models (\S 2), reveal the relation between magnetic turbulence and magnetic reconnection in \S 3, proceed to the discussion of the magnetic turbulence and magnetic field wandering in \S 4. We relate the model of fast turbulent reconnection to the Richarson diffusion in \S 5 and discuss flux freezing in astrophysical systems in \S 6. The connection with previous studies stressing the role of turbulence in reconnection is provided in \S 7. We discuss astrophysical implications of our work in
\S 8 and the summary is presented in \S 9.

\section{Astrophysical Reconnection versus Numerical Reconnection}

It is generally believed that a magnetic field embedded in a highly conductive 
fluid preserves its topology for all time due to the magnetic fields being frozen-in \cite{Alfven42,Parker79}.  
Although ionized astrophysical objects, like stars and galactic disks, are almost perfectly 
conducting, they show indications of changes in topology, ``magnetic reconnection'', 
on dynamical time scales \cite{Parker70}.  
Reconnection can be observed directly in the solar corona \cite{
YokoyamaShibata95}, but can also be inferred from the 
existence of large-scale dynamo activity inside stellar interiors \cite{Parker93, Ossendrijver03}.  
Solar flares \citep{Sturrock66} and $\gamma$-ray bursts (see \cite{Lazarianetal03, ZhangYan11}) 
are usually associated with magnetic reconnection.  A lot of previous work has concentrated on 
showing how reconnection can be rapid in plasmas with very small collisional rates 
\cite{Drake01, Drakeetal06}, which 
substantially constrains astrophysical applications of the corresponding 
reconnection models. 

A theory of magnetic reconnection is necessary to understand whether reconnection is represented correctly in numerical simulations. One should keep in mind that reconnection is fast in computer simulations due to high numerical diffusivity. Therefore, if there are situations where magnetic fields reconnect slowly, numerical simulations do not adequately reproduce the astrophysical reality. This means that if collisionless 
reconnection is indeed the only way to make reconnection fast, then the numerical simulations 
of many astrophysical processes, including those in interstellar media, which is collisional at the relevant scales, 
are in error. At the same time, it is not possible to conclude that reconnection must 
always be fast on the empirical grounds, as solar flares require periods of flux accumulation time, which correspond to slow reconnection.

To understand the difference between reconnection in astrophysical situations and 
in numerical simulations, one should recall that the dimensionless combination that
controls the resistive reconnection rate is the Lundquist number\footnote{The magnetic
Reynolds number, which is the ratio of the magnetic field decay time to the eddy
turnover time, is defined using the injection velocity $v_l$ as a characteristic
speed instead of the Alfv\'en speed $V_A$, which is taken in the Lundquist
number.}, defined as $S = L_xV_A / \lambda$, where $L_x$ is the length of the
reconnection layer, $V_A$ is the Alfv\'en velocity, and $\lambda=\eta c^2/4\pi$ 
is Ohmic diffusivity. Because of the huge astrophysical 
length-scales $L_x$ involved, the astrophysical Lundquist numbers are also huge,
e.g. for the ISM they are about $10^{16}$, while present-day MHD simulations
correspond to $S<10^4$. As the numerical efforts scale as $L_x^4$, where $L_x$
is the size of the box, it is feasible neither at present nor in the foreseeable future to have simulations with realistically Lundquist numbers. 

\section{Turbulence and Magnetic Reconnection}

While astrophysical fluids show a wide variety of properties in terms of their collisionality, degree of ionization, temperature etc., they share a common property, namely, most of the fluids are turbulent. The turbulent state of the fluids arises from large Reynolds numbers $Re\equiv LV/\nu$, where $L$ is the scale of the flow, $V$ is it velocity and $\nu$ is the viscosity, associated with astrophysical media. Note, that the large magnitude of $Re$ is mostly the consequence of the large astrophysical scales $L$ involved as well as the fact that (the field-perpendicular) viscosity is constrained by the presence of magnetic field. 

Observations of the interstellar medium reveal a Kolmogorov spectrum of electron density fluctuations (see
Ref.~\onlinecite{Armstrongetal94, ChepurnovLazarian10}) as well as steeper spectral slopes of supersonic velocity fluctuations (see Ref.~\onlinecite{Lazarian09} for a review). Measurement of the solar wind fluctuations also reveal turbulence power spectrum \cite{Leamon98}). Ubiquitous non-thermal broadening of spectral lines as well as measures obtained by other techniques (see Ref~\onlinecite{Burkhartetal10}) confirm that turbulence is present everywhere we test for its existence. As turbulence is known to change many processes, in particular the process of diffusion, the natural question is how it affects magnetic reconnection.

To deal with strong, dynamically important magnetic fields Lazarian \& Vishniac \cite{LazarianVishniac99} [henceforth LV99] proposed a model of fast reconnection in the presence of sub-Alfv\'enic turbulence. It is important to stress that unlike laboratory controlled settings, in astrophysical situations turbulence is preexisting, arising usually from the processes different from reconnection itself. We claim further in the paper that any modeling of astrophysical reconnection should account for the fact that magnetic reconnection takes place in the turbulent environment.
   
LV99 identified stochastic wandering of the magnetic field-lines 
as the most critical property of MHD turbulence which permits fast reconnection.  
As we discuss more fully below, this line-wandering widens the outflow region 
and alleviates the controlling constraint of mass conservation. The LV99 model 
has been successfully tested recently in Ref.\onlinecite{Kowaletal09} (see also higher resolution results 
in Ref.~\onlinecite{Lazarianetal10}). This model is radically different from its predecessors which also appealed 
to the effects of turbulence (see more comparisons in section 7). For instance, unlike Ref.~\onlinecite{Speiser70} and 
\onlinecite{Jacobson84} the model does not appeal to changes of the microscopic properties of the plasma. 
The nearest progenitor to LV99 was the work of  Matthaeus \& Lamkin \cite{MatthaeusLamkin85,MatthaeusLamkin86}, 
who studied the problem numerically in 2D MHD and who suggested that magnetic reconnection 
may be fast due to a number of  turbulence effects, e.g. multiple X points and turbulent EMF. However,  
Ref.~\onlinecite{MatthaeusLamkin85,MatthaeusLamkin86} did not realize the key role of played by magnetic field-line 
wandering, and did not obtain a quantitative prediction for the reconnection rate, as did LV99.

\section{Model of MHD turbulence and magnetic field wandering}

Wandering of magnetic field lines in LV99 model mostly depends on the Alfvenic component of magnetic perturbations. The exact scaling of the component, at least within the currently discussed turbulence models providing spectral indexes of $-5/3$ or $-3/2$ is not important, and the theoretical a possibility of fast reconnection was demonstrated in LV99 for turbulence with a wide range of spectral indexes and anisotropies. At the moment, we feel that the theory by Goldreich \& Sridhar\cite{GoldreichSridhar95} (henceforth GS95) is the most trusted one\footnote{It is discussed in Beresnyak \& Lazarian (2010) that the present day numerical simulations may not be sufficient to reveal the actual slope of Alfvenic cascade. In Ref.~\onlinecite{Beresnyak11} the slope of $-5/3$ was obtained in the largest reduced MHD simulations employing hyperdiffusion.}. Therefore in what follows we present our estimates using this theory. 

The GS95 theory was formulated originally for a situation when the energy injection happened at the Alfven velocity $v_A$. More general relations applicable to sub-Alfvenic turbulence were obtained in LV99, and we use those below. In particular, eddies in MHD turbulence 
become anisotropic and the key relation between the the parallel and perpendicular scales of eddies in the Alfenic turbulence is given by the so-called critical balance which can be written in terms of parallel $\ell_{\|}$ and perpendicular to magnetic field $\ell_{\bot}$ scales of the eddies (LV99):
\begin{equation}
\ell_{\|}\approx L_i \left(\frac{\ell_{\bot}}{L_i}\right)^{2/3} M_A^{-4/3}
\label{Lambda}
\end{equation}
where $L_i$ is the isotropic injection scale of the turbulence and $M_A\equiv u_L/v_A$ is the Alfv\'en Mach number of motions at the injection scale. Note, that the critical balance condition is only satisfied in the system of reference oriented with respect to the local magnetic field, which is different from the usual global system of reference related to the mean magnetic field. Thus we avoid using wave-vectors in characterizing the parallel and perpendicular scales\footnote{The wavevectors were used in GS95 where the distinction of local and global reference frames was not done and in a number of later papers where wavevectors were used mostly due to historic reasons. The first paper to discuss local system of reference was LV99.}.

In terms of perpendicular to magnetic field motions the scaling of Alfvenic turbulence is Kolmogorov-type:
\begin{equation}
\delta u_{\ell}\approx u_{L} \left(\frac{\ell_{\bot}}{L_i}\right)^{1/3} M_A^{1/3}.
\label{vl}
\end{equation}
which may be interpreted as a Kolmogorov cascade of mixing motions, which are not constrained by magnetic fields. 

The scaling relations for Alfvenic turbulence allow us to calculate the rate of 
magnetic field spreading. A bundle of field lines confined within a region of width $y$
at some particular point will spread out perpendicular to the mean
magnetic field direction as one moves in either direction following the local magnetic field lines.  The rate of field line diffusion
is given approximately by
\be
{d\langle y^2\rangle\over dx}\sim {\langle y^2\rangle\over \lambda_{\|}},
\ee
where $\lambda_{\|}^{-1}\approx \ell_{\|}^{-1}$, $\ell_{\|}$ is the parallel 
scale chosen so that the corresponding vertical scale, $\ell_{\perp}$,
is $\sim \langle y^2\rangle^{1/2}$, and $x$ is the
distance along an axis parallel to the mean magnetic field.
Therefore, using equation
(\ref{Lambda}) one gets
\be
{d\langle y^2\rangle\over dx}\sim L_i\left({\langle y^2\rangle\over L_i^2}\right)^{2/3}
\left({u_L\over v_A}\right)^{4/3} 
\label{eq:diffuse}
\ee
where we have 
substituted $\langle y^2\rangle ^{1/2}$ for $\ell_{\perp}$.  This expression for the
diffusion coefficient will only apply when $y$ is small enough for us
to use the strong turbulence scaling relations, or in other words when
$\langle y^2\rangle < L_i^2(u_L/v_A)^4$.  Larger bundles will diffuse at a maximum rate
of $L_i(u_L/v_A)^4$.  For $\langle y^2\rangle$ small equation (\ref{eq:diffuse}) implies
that a given field line will wander perpendicular to the mean field
line direction by an average amount
\be
\langle y^2\rangle^{1/2}\approx {x^{3/2}\over L_i^{1/2}} \left({u_L\over v_A}\right)^{2}
\label{eq:diffuse2}
\ee
in a distance $x$.  The fact that the rms perpendicular displacement grows
faster than $x$ is significant.  It implies that if
we consider a reconnection zone, a given magnetic flux element that
wanders out of the zone has only a small probability of wandering
back into it. This finding plays an important role both for magnetic reconnection in LV99 and heat transfer (see Ref.~\onlinecite{NarayanMedvedev01, Lazarian06}).

\section{Richardson Diffusion and LV99 model}

The advantage of the classical Sweet-Parker scheme of reconnection is that it naturally follows from the idea of Ohmic diffusion. Indeed, mass conservation requires that the inflow of matter through the scale of the contact region $L_x$ be equal to the outflow of matter through the diffusion layer $\Delta$, i.e.
\be   
v_{rec}= v_A {{\Delta}\over{L_x}}. 
\lb{vrec} 
\ee

\begin{figure}
\includegraphics[width=0.95\columnwidth,height=0.25\textheight]{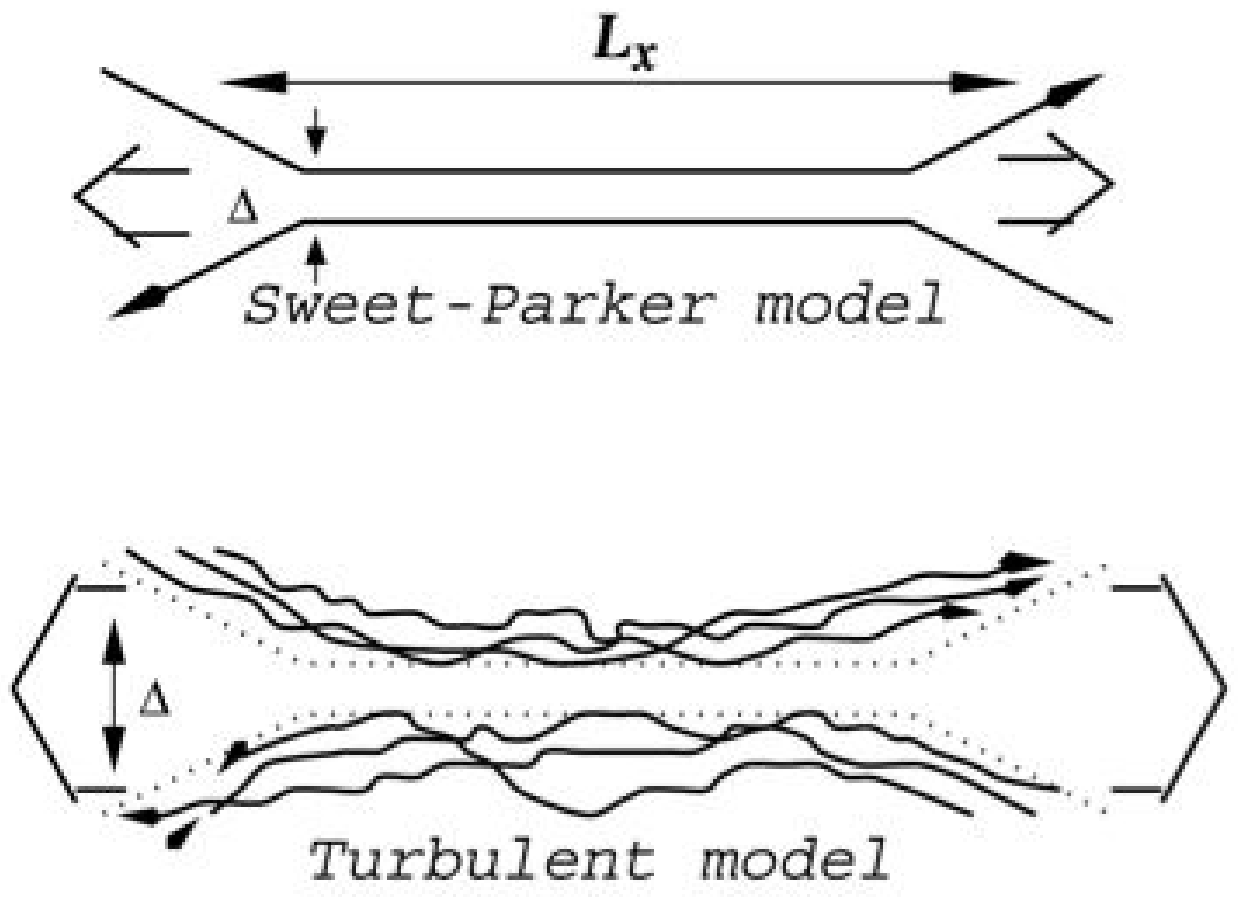}
\caption{Sweet-Parker laminar reconnection versus LV99 turbulent reconnection. Unlike the 
Sweet-Parker reconnection model , in the LV99 model the outflow is limited by magnetic field wandering rather than Ohmic diffusivity. From Lazarian et al. \cite{Lazarianetal04}. }
\end{figure}

\begin{figure}
\includegraphics[width=0.95\columnwidth,height=0.24\textheight]{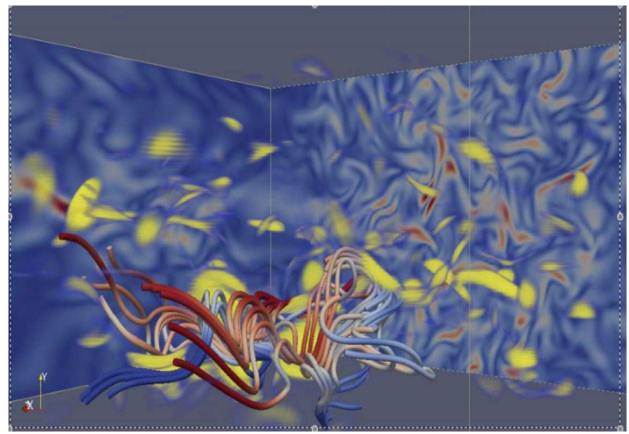}
\caption{Simulations of turbulent reconnection by Kowal et al. \cite{Kowaletal09}. Magnetic field
lines are visualized along with the fluctuations of current density. Substantial changes of the
magnetic field directions are due to reconnection, as the driving is subAlfvenic and the turbulent
perturbations of magnetic field are of low amplitude. }
\end{figure}

The mean-square 
vertical distance that a magnetic field-line can diffuse by resistivity in time $t$ is 
\be 
\langle y^2(t)\rangle \sim \lambda t. 
\lb{diff-dist} 
\ee
The field lines are advected out of the sides of the 
reconnection layer of length $L_x$ at a velocity of order $v_A.$ Thus, the time that the lines can 
spend in the resistive layer is the Alfv\'en crossing time $t_A=L_x/v_A.$ Thus, field lines can only 
be merged that are separated by a distance 
\be
\Delta = \sqrt{\langle y^2(t_A)\rangle} \sim \sqrt{\lambda t_A} = L_x/\sqrt{S},
\lb{Delta} 
\ee
where $S$ is Lundquist number. Combining Eqs. (\ref{vrec}) and (\ref{Delta}) one gets the famous Sweet-Parker reconnection rate, $v_{rec}=v_A/\sqrt{S}$. 

In LV99 magnetic field wandering determines the scale of the outflow $\Delta$ (see Figure 1). Using 
expressions from the earlier section one can obtain (LV99): 
\be
V_{rec}<v_A\min\left[\left({L_x\over L_i}\right)^{1/2},
\left({L_i\over L_x}\right)^{1/2}\right]
M_A^2.
\label{eq:lim2a}
\ee
This limit on the reconnection speed is fast, both in the sense that it does not depend on the resistivity, and in the sense that it represents a large
fraction of the Alfv\'en speed. To prove that Eq.~(\ref{eq:lim2a}) indeed constitutes the reconnection rate LV99 goes through a thorough job of considering all other possible bottlenecks for the reconnection and shows that they provide higher reconnection speed. It may be that because of the complexity of the argument, the LV99 theory was considered with caution by an appreciable part of the community till the time when it was successfully tested in Ref.~\onlinecite{Kowaletal09} (see Figure 2). 

Below we provide a new derivation of the LV99 reconnection rates which makes apparent that 
the LV99 model is a natural generalization of the laminar Sweet-Parker model to flows with background turbulence. 
The new argument in Eyink, Lazarian \& Vishniac \cite{Eyinketal11} (henceforth ELV11) is based on the concept of 
Richardson diffusion. It is known in hydrodynamic turbulence that the combination of small scale diffusion and large 
scale shear gives rise to Richardson diffusion, where the mean square separation between two particles grows as $t^3$ once the rms separation exceeds the viscous damping scale. A similar phenomenon occurs in MHD turbulence.  In both cases the separation at late times is independent of the microscopic transport coefficients.  Although the plasma is constrained to move along magnetic field lines, the combination of turbulence and ohmic dissipation produces a macroscopic region of points that are "downstream" from the same initial volume, even in the limit of vanishing resistivity.

Richardson diffusion (see Ref.~\onlinecite{Kupiainen03}) implies the mean squared separation of particles
$\langle |x_1(t)-x_2(t)|^2 \rangle\approx \epsilon t^3$, where $t$ is time, $\epsilon$ is the energy cascading rate and $\langleÉ\rangle$ denote an ensemble averaging. For subAlfvenic turbulence $\epsilon\approx u_L^4/(v_A L_i)$ (see LV99) and therefore analogously to Eq. (\ref{Delta}) one can write
\be
\Delta\approx \sqrt{\epsilon t_A^3}\approx L(L/L_i)^{1/2}M_A^2
\lb{D2}
\ee
where it is assumed that $L<L_i$. Combining Eqs. (\ref{vrec}) and (\ref{D2})
one gets
\be
v_{rec, LV99}\approx v_A (L/L_i)^{1/2}M_A^2.
\lb{LV99}
\ee
in the limit of $L<L_i$. Analogous considerations allow to recover the LV99 expression for $L>L_i$, which differs from Eq.~(\ref{LV99}) by the change of the
power $1/2$ to $-1/2$. These results coincide with those given by Eq.~(\ref{eq:lim2a}). 

It is important to stress that Richardson diffusion ultimately leads to diffusion over the entire width of large scale eddies once the plasma has moved the length of one such eddy. The precise scaling exponents for the turbulent cascade does not affect this result,
and all of the alternative scalings considered in LV99 yield the same behavior.

It is also important to emphasize that the original LV99 argument made no essential use of 
averaging over turbulent ensembles. The stochastic line-wandering which is 
the essence of their argument holds in every realization of the flow, at each instant of time. 
The ``spontaneous stochasticity" of field lines is not a statistical result in the usual sense of 
turbulence theory and does not arise from ensemble-averaging. The only use of ensembles 
in LV99 is to get a measure of the ``typical'' wandering distance $\Delta$ of the field lines (in 
an rms sense). If one looks at different ensemble members of the turbulent flow, or at different 
single-time snapshots of the steady reconnection state, then $\Delta$ will fluctuate. Thus, the 
reconnection rate will also fluctuate a considerable amount over ensemble members or over 
time.  E.g. see Figs.~12-14 in Ref.~\onlinecite{Kowaletal09}. But it will be  ``fast''  in each realization and 
at each instant of time, because the mass outflow constraint is lifted by the large wandering of field 
lines. 

The LV99 theory, therefore, does not involve ``turbulent resistivity'' or ``turbulent 
magnetic diffusivity'' as this is usually understood. This is ordinarily meant to be an enhanced diffusivity 
experienced by the ensemble-averaged magnetic field $\langle{\bf B}\rangle$. However, it is only if one 
assumes some scale-separation between the mean and the fluctuations that the effect of the fluctuations 
can be legitimately described as an enhanced diffusivity \cite{Moffatt83}. In realistic turbulent flows, 
with no scale-separation, this phenomenological description as an effective diffusivity can be 
wildly inaccurate. LV99 makes no appeal to such concepts and, indeed, never considers 
the ensemble-average field $\langle {\bf B}\rangle$ at all.

\section{Violation of Flux Freezing and Reconnection Diffusion}

While the derivation of the LV99 expressions in the previous section may look trivial, appealing to the concept of Richardson diffusion, in reality the justification of the treatment is rooted in fundamental progress achieved recently in understanding   the concept of frozen-in 
field-lines for turbulent MHD plasmas \cite{Eyink11} (see also Ref.~\onlinecite{EyinkAluie06,Eyink07,Eyink09}).

It is clear that in the presence of magnetic reconnection occurring densely in space the magnetic field lines cannot preserve their identity in turbulent flows. In fact, field-wandering is a process that is difficult to understand within the standard concept of flux freezing. Note that line-wandering 
implies that every space point is a nexus of infinitely many distinct lines. If magnetic field lines behaved like elastic threads which cannot change their topology, turbulence would create a system of unresolved magnetic knots draining the energy to small scales. Therefore, instead of fluid-type MHD one would get viscoelastic dynamics like Jello or rubber.

The high speed of reconnection given by equation (\ref{eq:lim2a})
naturally leads to a question of self-consistency.  Is it reasonable
to take the turbulent cascade suggested in GS95
when field lines in adjacent eddies are capable of reconnecting?
It turns out that in this context, our estimate for $V_{rec,global}$
is just fast enough to be interesting.  We note that when considering the 
intersection of nearly
parallel field lines in adjacent eddies the acceleration of plasma
from the reconnection layer due to the pressure gradient
is not $\ell_{\|}^{-1}v_A^2$, but rather $(\ell_{\|}^3/\ell_{\perp}^2)^{-1}v_A^2$,
since only the energy of the component of the magnetic field 
which is not shared is available to drive the outflow.  On the
other hand, the characteristic length contraction of a given
field line due to reconnection between adjacent eddies is only
$(\ell_{\|}/\ell_{\perp}^2)^{-1}$.  This gives an effective ejection rate 
of $v_A/\ell_{\|}$.  Since the width of the diffusion layer over a 
length $\ell_{\|}$ is just $\ell_{\perp}$, we can replace 
equation (\ref{eq:lim2a}) with
\be
v_{rec,eddy}\approx v_A {\ell_{\bot}\over \ell_{\|}}. 
\ee
The associated reconnection rate is just
\be
\tau^{-1}_{reconnect}\sim v_A/\ell_{\|}, 
\ee
which in GS95 is just the nonlinear cascade rate on the scale
$\ell_{\|}$.  Note, that this result is general and does not
involve assuming that GS95 model of turbulence is correct.  
It, however, assures that the turbulent reconnection rates are high enough to avoid the formation of unresolved magnetic field knots, i.e. unresolved field line intersections, in the GS95 and similar models of 3D turbulence.

Another consequence of magnetic reconnection is the diffusion of plasma between different magnetic lines in turbulent flow. This process allows convective heat transfer as described in Ref.~\onlinecite{Choetal03} as well as magnetic field removal from molecular clouds and accretion disks \cite{Lazarian05, SantosdeLimaetal10}. The astrophysical diffusion enabled by reconnection of magnetic field lines gives rise to a concept of {\it reconnection diffusion}, which for many processes, e.g. for various stages of star formation may be more efficient than the ambipolar diffusion. Below we briefly
discuss this new concept (see also Ref. \onlinecite{Lazarian05, Lazarianetal10, Lazarian11}).

\begin{figure}
\includegraphics[width=0.95\columnwidth,height=0.24\textheight]{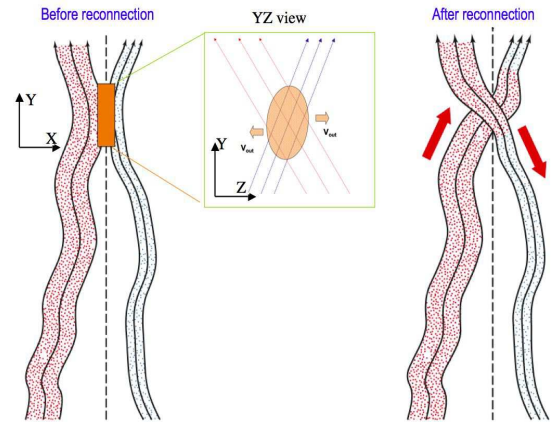}
\caption{Reconnection diffusion: process 1. Illustration of the mixing of matter due to reconnection as two flux tubes of different magnetic field strength interact.}
\end{figure} 

\begin{figure}
\includegraphics[width=0.95\columnwidth,height=0.24\textheight]{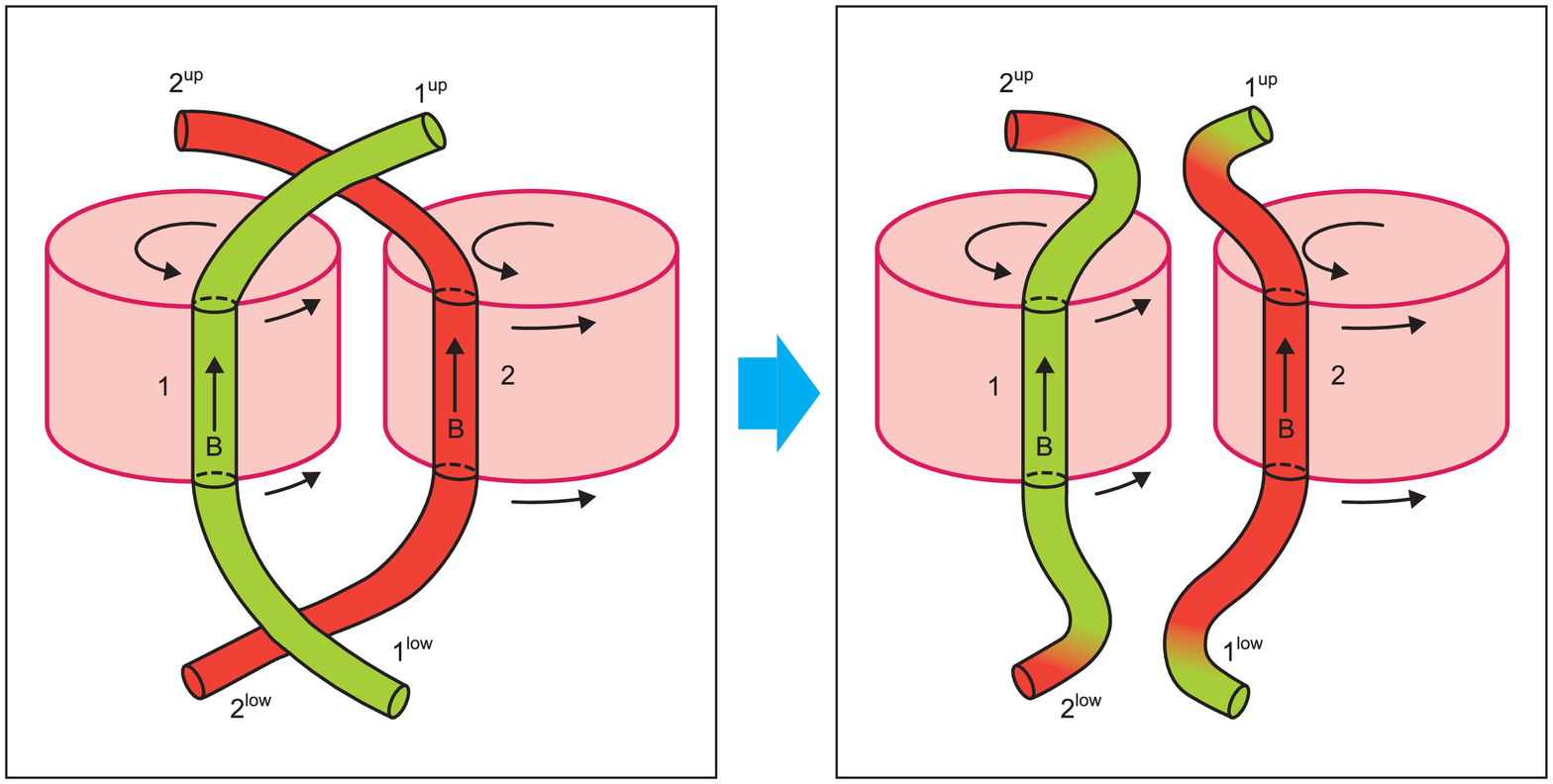}
\caption{Reconnection diffusion: process 2. Illustration of the mixing of matter due to reconnection as two flux tubes of different magnetic field strength interact.}
\end{figure} 
 
The common wisdom based on the notion of nearly perfect flux freezing for astrophysical
magnetic fields suggests that mixing of plasma entrained on different flux tubes is impossible.
However, it is easy to see that this is not true in the presence of LV99 reconnection. First of
all, the process can connect magnetic field lines with different plasma loading, inducing diffusion
and mixing along the newly emerging magnetic field lines. Then, in turbulent fluid the magnetic
 field lines can be shredded and mixed by eddies of smaller and smaller scales. These two
processes are illustrated by separate figures below, but, in reality, these two processes in 
turbulent magnetized fluid take place simultaneously.

Figure 3 illustrates the concept of reconnection diffusion for two flux tubes which have the same total pressure $P_{tot}=P_{plasma}+P_{magn}$. In the absence of turbulence the two flux tubes are in pressure equilibrium and the entrained plasma stays on the flux tubes. In the presence of 3D turbulence flux tubes can reconnect (minimizing the energy of the $Z$-component of magnetic field) which allows plasma flow from one flux tube to another. This process is an illustration of a multi-scale process taking place in realistic turbulent flows.

Figure 4 illustrates the concept of reconnection diffusion when the magnetic pressure in the 
flux tubes is the same. This is, for instance, is the case of heat advection by turbulence. The mixing is happening as new magnetic flux tubes are constantly formed 
from magnetic flux tubes that belong to different eddies. In the figure two adjacent eddies are 
shown and the process is limited to the effects of eddies of a single scale. It is clear that 
plasmas which was originally entrained over different flux tubes gets into contact along the new
emerging flux tubes. The process similar to the depicted one takes place at different scales down
to the scale of the smallest eddies. Molecular diffusivity then takes over. In the case of heat 
transfer small scale molecular diffusivity will ensure that the temperature along the newly 
formed magnetic flux tubes is the same.

Naturally, the process illustrated by Figure 4 also happens when the pressure of plasmas along magnetic flux tubes is different. LV99 process of magnetic field reconnection ensures that {\it both}
magnetic field and plasmas are diffusing at the turbulent diffusion rate. We also note that in the 
presence of forcing, e.g. gravitational forces acting upon the conducting gas, the diffusion will be accompanied by the removal of magnetic field from the center of the gravitational potential due to
the magnetic field bouyancy. It is important to understand that the process in Figure 4 is limited only by the velocity of the eddies. Therefore supersonic turbulence can induce supersonic mixing.

Reconnection diffusion is due to eddies that are perpendicular to the {\it local} direction of
magnetic field. This direction, in general, does not coincide with the mean magnetic field direction. Therefore in the lab system of reference related to the mean magnetic field the diffusion of magnetic field and plasmas will happen both parallel and perpendicular to the mean magnetic field direction.

\section{Relevant work on reconnection rates}

Having discussed the LV99 model and new arguments supporting it, it is worth 
considering recent work on alternative approaches to calculating reconnection rates.
Over the last decade, more traditional approaches to reconnection have 
changed considerably. At the time of its introduction, the models competing with LV99 
were modifications of the single X-point collisionless reconnection scheme first introduced by Petschek \cite{petschek64}. 
Those models had 
point-wise localized reconnection regions which were stabilized via plasma effects so that the
outflow opened up on larger scales. Such configurations would be difficult to 
realize in the presence of random forcing,  which would be expected to collapse the reconnection layer. 
Moreover, Refs.\onlinecite{CiaravellaRaymond08} argued that observations of solar flares were inconsistent with single X-point 
reconnection.

In response to these objections, more recent models of collisionless reconnection have acquired several features in common
with the LV99 model. In particular, they have moved to consideration of volume filling reconnection, 
(although it is not clear how this volume filling is achieved in the presence of a 
single reconnection layer (see Ref.~\onlinecite{Drakeetal06})).  While much of the discussion still centers 
around magnetic islands produced by reconnection, in three dimensions these islands are expected to evolve into 
contracting 3D loops or ropes\cite{Daughtonetal08}, which is broadly similar to what is depicted 
in Figure 1, at least in the sense of introducing stochasticity to the reconnection zone.

The departure from the concept of laminar reconnection and the introduction of magnetic stochasticity 
is also apparent in a number of the recent papers appealing to the tearing mode instability to drive fast 
reconnection (see Refs.~\onlinecite{Loureiroetal09, Bhattacharjeeetal09})\footnote{The idea of appealing to the tearing mode as a means of enhancing the reconnection 
speed can be traced back to Refs.~\onlinecite{Strauss88 ShibataTanuma01}.} 
LV99 showed that the linear growth of tearing modes is insufficient to obtain fast reconnection. 
More recent work is based on the idea that the non-linear growth of magnetic islands due to mergers 
provides large scale growth rates larger than the tearing mode linear growth rates on these scales.  
A situation where the non-linear growth is faster than the linear one is 
rather unusual and requires further investigation (see Ref.~\onlinecite{Diamondetal84}.)  
Since tearing modes exist even in a collisional fluid, this may open another channel of 
reconnection in such fluids. This reconnection, as we discuss below, should not be ``too fast'' to account for the observational data.

If local turbulence is driven by release of energy from the magnetic field, 
it may result in a runaway turbulent reconnection process which may be relevant 
to some numerical simulations \cite{Lapenta08, BettariniLapenta09}.
Alternatively, if tearing modes begin
by driving relatively slow reconnection then a similar runaway might result \cite{LapentaLazarian11}.
 
In any case, in most astrophysical situations one has to deal with the {\it pre-existing} 
turbulence, which is the consequence of high Reynolds number of astrophysical fluids. Such 
turbulence may modify or suppress instabilities, including the tearing mode instability. In this paper we have shown that it, by itself, induces fast reconnection on dynamical time scales. 

\section{Astrophysical Implications}

Fast magnetic reconnection induces numerious astrophysical implications. In the sections above 
we have discussed a new concept of reconnection diffusion that is likely to dominate diffusion
of magnetic fields and plasmas in various astrophysical environments from accretion disks to 
intracluster medium (see e.g. Refs. \onlinecite{Fabianetal11, SantosLimaetal11}). 
The role of reconnection diffusion may be different. For instance, 
in circumstellar accretion disks and cores of molecular clouds it removes magnetic flux, while
in more diffuse interstellar medium it mostly provides good mixing destroying the correlation of 
magnetic field strength with plasma density. Turbulent transport of heat, as well as impurities and
dust is happening due to the reconnection diffusion.

At the same time, the LV99 magnetic reconnection induces a number of new effects. For instance, particles trapped over magnetic field lines get accelerated via First order Fermi process \cite{deGouveiadalPinoLazarian05, Lazarian05}. This acceleration is also reported in a situation when loops of magnetic field are formed via tearing reconnection \cite{Drakeetal06, Drakeetal09}. The physics is the same, namely, in both cases magnetic fields shrink and the particles entrained
over magnetic field lines get accelerated. This provides a likely explanation of the origin of
the anomalous cosmic rays measured by Voyagers \cite{LazarianOpher09, Drakeetal10}  as well as the anisotropy of cosmic rays
in the direction of heliotail \cite{LazarianDesiati10} reported by different groups. We would like to stress
that the acceleration in 3D and 2D is found to be different \cite{Kowaletal11} and therefore results of 2D 
numerical experiments dealing with particle acceleration should be taken with grain of salt. 

Naturally, the acceleration of energetic particles and reconnection diffusion do not encompass all
the possible applications of the LV99 model of reconnection. For instance, Lazarian et al \cite{Lazarianetal03}
proposed a way of explaining gamma ray bursts appealing to turbulent reconnection. This idea was
further elaborated in a high impact paper by Zhang \& Yan \cite{ZhangYan11} In general, bursts and flares are the natural
consequence of the LV99 model. If magnetic fields are originally laminar, the low reconnection rate
allows the accummulation of magnetic flux. As turbulence increases, for example, due to the outflow,
the reconnection rate increases, making the outflow more turbulent. This induces positive feedback
resulting in what can be termed ``reconnection instability''.

Whatever the particular astrophysical consequences of LV99 model, the most striking is the fact
that in turbulent fluids the basic idea of flux freezing is dramatically modified in turbulent fluids. This shakes the foundations of the astrophysical MHD, as astrophysical fluids are usually 
turbulent.

\section{Summary}
Our main points can be summarized as follows:

Astrophysical fluids are turbulent and turbulence must be accounted in the models of astrophysical magnetic reconnection.

Turbulence and magnetic reconnection are two interdependent processes: turbulence makes magnetic reconnection fast, but magnetic reconnection is required for the turbulence to evolve in a self-similar fashion.

Plasma effects may be important for local reconnection effects, but this in most
cases will not affect the resulting global reconnection rates.

\begin{acknowledgments}

We thank A. Bhattacharjee, P. H. Diamond, and A. Pouquet 
for some useful discussions.  AL is supported by the NSF grant AST 0808118, NASA grant NNX09AH78G and the Center for Magnetic Self Organization. AL acknowledges Humboldt Award at the Universities
of Cologne and Bochum and a Fellowship at the International Institute of Physics (Brazil). 
GE was partially supported by NSF grants AST 0428325 and CDI-II:  CMMI 0941530. 
The work of ETV is supported by the National Science and Engineering Research 
Council of Canada.
\end{acknowledgments}









%




%






\bibliography{lazarian_recon_proc.bib}

\providecommand{\noopsort}[1]{}\providecommand{\singleletter}[1]{#1}%
\begin{thebibliography}{56}%
\makeatletter
\providecommand \@ifxundefined [1]{%
 \@ifx{#1\undefined}
}%
\providecommand \@ifnum [1]{%
 \ifnum #1\expandafter \@firstoftwo
 \else \expandafter \@secondoftwo
 \fi
}%
\providecommand \@ifx [1]{%
 \ifx #1\expandafter \@firstoftwo
 \else \expandafter \@secondoftwo
 \fi
}%
\providecommand \natexlab [1]{#1}%
\providecommand \enquote  [1]{``#1''}%
\providecommand \bibnamefont  [1]{#1}%
\providecommand \bibfnamefont [1]{#1}%
\providecommand \citenamefont [1]{#1}%
\providecommand \href@noop [0]{\@secondoftwo}%
\providecommand \href [0]{\begingroup \@sanitize@url \@href}%
\providecommand \@href[1]{\@@startlink{#1}\@@href}%
\providecommand \@@href[1]{\endgroup#1\@@endlink}%
\providecommand \@sanitize@url [0]{\catcode `\\12\catcode `\$12\catcode
  `\&12\catcode `\#12\catcode `\^12\catcode `\_12\catcode `\%12\relax}%
\providecommand \@@startlink[1]{}%
\providecommand \@@endlink[0]{}%
\providecommand \url  [0]{\begingroup\@sanitize@url \@url }%
\providecommand \@url [1]{\endgroup\@href {#1}{\urlprefix }}%
\providecommand \urlprefix  [0]{URL }%
\providecommand \Eprint [0]{\href }%
\providecommand \doibase [0]{http://dx.doi.org/}%
\providecommand \selectlanguage [0]{\@gobble}%
\providecommand \bibinfo  [0]{\@secondoftwo}%
\providecommand \bibfield  [0]{\@secondoftwo}%
\providecommand \translation [1]{[#1]}%
\providecommand \BibitemOpen [0]{}%
\providecommand \bibitemStop [0]{}%
\providecommand \bibitemNoStop [0]{.\EOS\space}%
\providecommand \EOS [0]{\spacefactor3000\relax}%
\providecommand \BibitemShut  [1]{\csname bibitem#1\endcsname}%
\let\auto@bib@innerbib\@empty
\bibitem [{\citenamefont {Alfv\'{e}n}(1942)}]{Alfven42}%
  \BibitemOpen
  \bibfield  {author} {\bibinfo {author} {\bibfnamefont {H.}~\bibnamefont
  {Alfv\'{e}n}},\ }\href@noop {} {\bibfield  {journal} {\bibinfo  {journal}
  {Ark. Mat., Astron. o. Fys.}\ }\textbf {\bibinfo {volume} {29B}},\ \bibinfo
  {pages} {1} (\bibinfo {year} {1942})}\BibitemShut {NoStop}%
\bibitem [{\citenamefont {Parker}(1979)}]{Parker79}%
  \BibitemOpen
  \bibfield  {author} {\bibinfo {author} {\bibfnamefont {E.~N.}\ \bibnamefont
  {Parker}},\ }\href@noop {} {\emph {\bibinfo {title} {Cosmic Magnetic
  Fields}}}\ (\bibinfo  {publisher} {Oxford University Press},\ \bibinfo {year}
  {1979})\BibitemShut {NoStop}%
\bibitem [{\citenamefont {Parker}(1970)}]{Parker70}%
  \BibitemOpen
  \bibfield  {author} {\bibinfo {author} {\bibfnamefont {E.~N.}\ \bibnamefont
  {Parker}},\ }\href@noop {} {\bibfield  {journal} {\bibinfo  {journal} {ApJ}\
  }\textbf {\bibinfo {volume} {162}},\ \bibinfo {pages} {665} (\bibinfo {year}
  {1970})}\BibitemShut {NoStop}%
\bibitem [{\citenamefont {Yokoyama}\ and\ \citenamefont
  {Shibata}(1995)}]{YokoyamaShibata95}%
  \BibitemOpen
  \bibfield  {author} {\bibinfo {author} {\bibfnamefont {T.}~\bibnamefont
  {Yokoyama}}\ and\ \bibinfo {author} {\bibfnamefont {K.}~\bibnamefont
  {Shibata}},\ }\href@noop {} {\bibfield  {journal} {\bibinfo  {journal}
  {Nature}\ }\textbf {\bibinfo {volume} {375}},\ \bibinfo {pages} {42}
  (\bibinfo {year} {1995})}\BibitemShut {NoStop}%
\bibitem [{\citenamefont {Parker}(1993)}]{Parker93}%
  \BibitemOpen
  \bibfield  {author} {\bibinfo {author} {\bibfnamefont {E.~N.}\ \bibnamefont
  {Parker}},\ }\href@noop {} {\bibfield  {journal} {\bibinfo  {journal} {ApJ}\
  }\textbf {\bibinfo {volume} {408}},\ \bibinfo {pages} {707} (\bibinfo {year}
  {1993})}\BibitemShut {NoStop}%
\bibitem [{\citenamefont {Ossendrijver}(2003)}]{Ossendrijver03}%
  \BibitemOpen
  \bibfield  {author} {\bibinfo {author} {\bibfnamefont {M.}~\bibnamefont
  {Ossendrijver}},\ }\href@noop {} {\bibfield  {journal} {\bibinfo  {journal}
  {AandA Rev.}\ }\textbf {\bibinfo {volume} {11}},\ \bibinfo {pages} {287}
  (\bibinfo {year} {2003})}\BibitemShut {NoStop}%
\bibitem [{\citenamefont {Sturrock}(1966)}]{Sturrock66}%
  \BibitemOpen
  \bibfield  {author} {\bibinfo {author} {\bibfnamefont {P.~A.}\ \bibnamefont
  {Sturrock}},\ }\href@noop {} {\bibfield  {journal} {\bibinfo  {journal}
  {Nature}\ }\textbf {\bibinfo {volume} {211}},\ \bibinfo {pages} {695}
  (\bibinfo {year} {1966})}\BibitemShut {NoStop}%
\bibitem [{\citenamefont {Lazarian}\ \emph {et~al.}(2003)\citenamefont
  {Lazarian}, \citenamefont {Petrosian}, \citenamefont {Yan},\ and\
  \citenamefont {Cho}}]{Lazarianetal03}%
  \BibitemOpen
  \bibfield  {author} {\bibinfo {author} {\bibfnamefont {A.}~\bibnamefont
  {Lazarian}}, \bibinfo {author} {\bibfnamefont {V.}~\bibnamefont {Petrosian}},
  \bibinfo {author} {\bibfnamefont {H.}~\bibnamefont {Yan}}, \ and\ \bibinfo
  {author} {\bibfnamefont {J.}~\bibnamefont {Cho}},\ }\href@noop {} {}\bibinfo
  {howpublished} {Beaming and Jets in Gamma Ray Bursts} (\bibinfo {year}
  {2003}),\ \bibinfo {note} {proceedings, volume 45}\BibitemShut {NoStop}%
\bibitem [{\citenamefont {Zhang}\ and\ \citenamefont {Yan}(2011)}]{ZhangYan11}%
  \BibitemOpen
  \bibfield  {author} {\bibinfo {author} {\bibfnamefont {B.}~\bibnamefont
  {Zhang}}\ and\ \bibinfo {author} {\bibfnamefont {H.}~\bibnamefont {Yan}},\
  }\href@noop {} {\bibfield  {journal} {\bibinfo  {journal} {ApJ}\ }\textbf
  {\bibinfo {volume} {726}},\ \bibinfo {pages} {90} (\bibinfo {year}
  {2011})}\BibitemShut {NoStop}%
\bibitem [{\citenamefont {Drake}(2001)}]{Drake01}%
  \BibitemOpen
  \bibfield  {author} {\bibinfo {author} {\bibfnamefont {J.~F.}\ \bibnamefont
  {Drake}},\ }\href@noop {} {\bibfield  {journal} {\bibinfo  {journal}
  {Nature}\ }\textbf {\bibinfo {volume} {410}},\ \bibinfo {pages} {525}
  (\bibinfo {year} {2001})}\BibitemShut {NoStop}%
\bibitem [{\citenamefont {Drake}\ \emph {et~al.}(2006)\citenamefont {Drake},
  \citenamefont {Swisdak}, \citenamefont {Che},\ and\ \citenamefont
  {Shay}}]{Drakeetal06}%
  \BibitemOpen
  \bibfield  {author} {\bibinfo {author} {\bibfnamefont {J.~F.}\ \bibnamefont
  {Drake}}, \bibinfo {author} {\bibfnamefont {M.}~\bibnamefont {Swisdak}},
  \bibinfo {author} {\bibfnamefont {H.}~\bibnamefont {Che}}, \ and\ \bibinfo
  {author} {\bibfnamefont {M.~A.}\ \bibnamefont {Shay}},\ }\href@noop {}
  {\bibfield  {journal} {\bibinfo  {journal} {Nature}\ }\textbf {\bibinfo
  {volume} {443}},\ \bibinfo {pages} {553} (\bibinfo {year}
  {2006})}\BibitemShut {NoStop}%
\bibitem [{Note1()}]{Note1}%
  \BibitemOpen
  \bibinfo {note} {The magnetic Reynolds number, which is the ratio of the
  magnetic field decay time to the eddy turnover time, is defined using the
  injection velocity $v_l$ as a characteristic speed instead of the Alfv\'en
  speed $V_A$, which is taken in the Lundquist number.}\BibitemShut {Stop}%
\bibitem [{\citenamefont {Armstrong}\ and\ \citenamefont
  {Spangler}(1995)}]{Armstrongetal94}%
  \BibitemOpen
  \bibfield  {author} {\bibinfo {author} {\bibfnamefont {R.~B.~J.}\
  \bibnamefont {Armstrong}, \bibfnamefont {J.~W.}}\ and\ \bibinfo {author}
  {\bibfnamefont {S.~R.}\ \bibnamefont {Spangler}},\ }\href@noop {} {\bibfield
  {journal} {\bibinfo  {journal} {ApJ}\ }\textbf {\bibinfo {volume} {443}},\
  \bibinfo {pages} {209} (\bibinfo {year} {1995})}\BibitemShut {NoStop}%
\bibitem [{\citenamefont {Chepurnov}\ and\ \citenamefont
  {Lazarian}(2010)}]{ChepurnovLazarian10}%
  \BibitemOpen
  \bibfield  {author} {\bibinfo {author} {\bibfnamefont {A.}~\bibnamefont
  {Chepurnov}}\ and\ \bibinfo {author} {\bibfnamefont {A.}~\bibnamefont
  {Lazarian}},\ }\href@noop {} {\bibfield  {journal} {\bibinfo  {journal}
  {ApJ}\ }\textbf {\bibinfo {volume} {710}},\ \bibinfo {pages} {853} (\bibinfo
  {year} {2010})}\BibitemShut {NoStop}%
\bibitem [{\citenamefont {{Lazarian}}(2009)}]{Lazarian09}%
  \BibitemOpen
  \bibfield  {author} {\bibinfo {author} {\bibfnamefont {A.}~\bibnamefont
  {{Lazarian}}},\ }\href {\doibase 10.1007/s11214-008-9460-y} {\bibfield
  {journal} {\bibinfo  {journal} {Space Science Reivews}\ }\textbf {\bibinfo
  {volume} {143}},\ \bibinfo {pages} {357} (\bibinfo {year} {2009})},\ \Eprint
  {http://arxiv.org/abs/0811.0839} {arXiv:0811.0839} \BibitemShut {NoStop}%
\bibitem [{\citenamefont {{Leamon}}\ \emph {et~al.}(1998)\citenamefont
  {{Leamon}}, \citenamefont {{Smith}}, \citenamefont {{Ness}}, \citenamefont
  {{Matthaeus}},\ and\ \citenamefont {{Wong}}}]{Leamon98}%
  \BibitemOpen
  \bibfield  {author} {\bibinfo {author} {\bibfnamefont {R.~J.}\ \bibnamefont
  {{Leamon}}}, \bibinfo {author} {\bibfnamefont {C.~W.}\ \bibnamefont
  {{Smith}}}, \bibinfo {author} {\bibfnamefont {N.~F.}\ \bibnamefont {{Ness}}},
  \bibinfo {author} {\bibfnamefont {W.~H.}\ \bibnamefont {{Matthaeus}}}, \ and\
  \bibinfo {author} {\bibfnamefont {H.~K.}\ \bibnamefont {{Wong}}},\ }\href
  {\doibase 10.1029/97JA03394} {\bibfield  {journal} {\bibinfo  {journal}
  {{Journal Geophis. Res.}}\ }\textbf {\bibinfo {volume} {103}},\ \bibinfo
  {pages} {4775} (\bibinfo {year} {1998})}\BibitemShut {NoStop}%
\bibitem [{\citenamefont {Burkhart}\ \emph {et~al.}(2010)\citenamefont
  {Burkhart}, \citenamefont {Stanimirovi{\'c}}, \citenamefont {Lazarian},\ and\
  \citenamefont {Kowal}}]{Burkhartetal10}%
  \BibitemOpen
  \bibfield  {author} {\bibinfo {author} {\bibfnamefont {B.}~\bibnamefont
  {Burkhart}}, \bibinfo {author} {\bibfnamefont {S.}~\bibnamefont
  {Stanimirovi{\'c}}}, \bibinfo {author} {\bibfnamefont {A.}~\bibnamefont
  {Lazarian}}, \ and\ \bibinfo {author} {\bibfnamefont {G.}~\bibnamefont
  {Kowal}},\ }\href@noop {} {\bibfield  {journal} {\bibinfo  {journal} {ApJ}\
  }\textbf {\bibinfo {volume} {708}},\ \bibinfo {pages} {1204} (\bibinfo {year}
  {2010})}\BibitemShut {NoStop}%
\bibitem [{\citenamefont {Lazarian}\ and\ \citenamefont
  {Vishniac}(1999)}]{LazarianVishniac99}%
  \BibitemOpen
  \bibfield  {author} {\bibinfo {author} {\bibfnamefont {A.}~\bibnamefont
  {Lazarian}}\ and\ \bibinfo {author} {\bibfnamefont {E.~T.}\ \bibnamefont
  {Vishniac}},\ }\href@noop {} {\bibfield  {journal} {\bibinfo  {journal}
  {ApJ}\ }\textbf {\bibinfo {volume} {517}},\ \bibinfo {pages} {700} (\bibinfo
  {year} {1999})}\BibitemShut {NoStop}%
\bibitem [{\citenamefont {Kowal}\ \emph {et~al.}(2009)\citenamefont {Kowal},
  \citenamefont {Lazarian}, \citenamefont {Vishniac},\ and\ \citenamefont
  {Otmianowska-Mazur}}]{Kowaletal09}%
  \BibitemOpen
  \bibfield  {author} {\bibinfo {author} {\bibfnamefont {G.}~\bibnamefont
  {Kowal}}, \bibinfo {author} {\bibfnamefont {A.}~\bibnamefont {Lazarian}},
  \bibinfo {author} {\bibfnamefont {E.~T.}\ \bibnamefont {Vishniac}}, \ and\
  \bibinfo {author} {\bibfnamefont {K.}~\bibnamefont {Otmianowska-Mazur}},\
  }\href@noop {} {\bibfield  {journal} {\bibinfo  {journal} {ApJ}\ }\textbf
  {\bibinfo {volume} {700}},\ \bibinfo {pages} {63} (\bibinfo {year}
  {2009})}\BibitemShut {NoStop}%
\bibitem [{\citenamefont {{Lazarian}}\ \emph {et~al.}(2010)\citenamefont
  {{Lazarian}}, \citenamefont {{Santos-Lima}},\ and\ \citenamefont {{de Gouveia
  Dal Pino}}}]{Lazarianetal10}%
  \BibitemOpen
  \bibfield  {author} {\bibinfo {author} {\bibfnamefont {A.}~\bibnamefont
  {{Lazarian}}}, \bibinfo {author} {\bibfnamefont {R.}~\bibnamefont
  {{Santos-Lima}}}, \ and\ \bibinfo {author} {\bibfnamefont {E.}~\bibnamefont
  {{de Gouveia Dal Pino}}},\ }in\ \href@noop {} {\emph {\bibinfo {booktitle}
  {Numerical Modeling of Space Plasma Flows, Astronum-2009}}},\ \bibinfo
  {series} {Astronomical Society of the Pacific Conference Series}, Vol.\
  \bibinfo {volume} {429},\ \bibinfo {editor} {edited by\ \bibinfo {editor}
  {\bibnamefont {{N.~V.~Pogorelov, E.~Audit, \& G.~P.~Zank}}}}\ (\bibinfo
  {year} {2010})\ pp.\ \bibinfo {pages} {113--+},\ \Eprint
  {http://arxiv.org/abs/1003.2640} {arXiv:1003.2640 [astro-ph.GA]} \BibitemShut
  {NoStop}%
\bibitem [{\citenamefont {Speiser}(1970)}]{Speiser70}%
  \BibitemOpen
  \bibfield  {author} {\bibinfo {author} {\bibfnamefont {T.~W.}\ \bibnamefont
  {Speiser}},\ }\href@noop {} {\bibfield  {journal} {\bibinfo  {journal} {Plan.
  and Space Sci.}\ }\textbf {\bibinfo {volume} {18}},\ \bibinfo {pages} {613}
  (\bibinfo {year} {1970})}\BibitemShut {NoStop}%
\bibitem [{\citenamefont {Jacobson}\ and\ \citenamefont
  {Moses}(1984)}]{Jacobson84}%
  \BibitemOpen
  \bibfield  {author} {\bibinfo {author} {\bibfnamefont {A.~R.}\ \bibnamefont
  {Jacobson}}\ and\ \bibinfo {author} {\bibfnamefont {R.~W.}\ \bibnamefont
  {Moses}},\ }\href@noop {} {\bibfield  {journal} {\bibinfo  {journal} {Phys.
  Rev. A.}\ }\textbf {\bibinfo {volume} {29}},\ \bibinfo {pages} {3335}
  (\bibinfo {year} {1984})}\BibitemShut {NoStop}%
\bibitem [{\citenamefont {Matthaeus}\ and\ \citenamefont
  {Lamkin}(1985)}]{MatthaeusLamkin85}%
  \BibitemOpen
  \bibfield  {author} {\bibinfo {author} {\bibfnamefont {W.~H.}\ \bibnamefont
  {Matthaeus}}\ and\ \bibinfo {author} {\bibfnamefont {S.~L.}\ \bibnamefont
  {Lamkin}},\ }\href@noop {} {\bibfield  {journal} {\bibinfo  {journal} {Phys.
  Fluids, 28}\ }\textbf {\bibinfo {volume} {28}},\ \bibinfo {pages} {303}
  (\bibinfo {year} {1985})}\BibitemShut {NoStop}%
\bibitem [{\citenamefont {Matthaeus}\ and\ \citenamefont
  {Lamkin}(1986)}]{MatthaeusLamkin86}%
  \BibitemOpen
  \bibfield  {author} {\bibinfo {author} {\bibfnamefont {W.~H.}\ \bibnamefont
  {Matthaeus}}\ and\ \bibinfo {author} {\bibfnamefont {S.~L.}\ \bibnamefont
  {Lamkin}},\ }\href@noop {} {\bibfield  {journal} {\bibinfo  {journal} {Phys.
  Fluids}\ }\textbf {\bibinfo {volume} {29}},\ \bibinfo {pages} {2513}
  (\bibinfo {year} {1986})}\BibitemShut {NoStop}%
\bibitem [{\citenamefont {Goldreich}\ and\ \citenamefont
  {Sridhar}(1995)}]{GoldreichSridhar95}%
  \BibitemOpen
  \bibfield  {author} {\bibinfo {author} {\bibfnamefont {P.}~\bibnamefont
  {Goldreich}}\ and\ \bibinfo {author} {\bibfnamefont {S.}~\bibnamefont
  {Sridhar}},\ }\href@noop {} {\bibfield  {journal} {\bibinfo  {journal} {ApJ}\
  }\textbf {\bibinfo {volume} {438}},\ \bibinfo {pages} {763} (\bibinfo {year}
  {1995})}\BibitemShut {NoStop}%
\bibitem [{Note2()}]{Note2}%
  \BibitemOpen
  \bibinfo {note} {It is discussed in Beresnyak \& Lazarian (2010) that the
  present day numerical simulations may not be sufficient to reveal the actual
  slope of Alfvenic cascade. In Ref.~\protect \rev@citealp {Beresnyak11} the
  slope of $-5/3$ was obtained in the largest reduced MHD simulations employing
  hyperdiffusion.}\BibitemShut {Stop}%
\bibitem [{Note3()}]{Note3}%
  \BibitemOpen
  \bibinfo {note} {The wavevectors were used in GS95 where the distinction of
  local and global reference frames was not done and in a number of later
  papers where wavevectors were used mostly due to historic reasons. The first
  paper to discuss local system of reference was LV99.}\BibitemShut {Stop}%
\bibitem [{\citenamefont {Narayan}\ and\ \citenamefont
  {Medvedev}(2001)}]{NarayanMedvedev01}%
  \BibitemOpen
  \bibfield  {author} {\bibinfo {author} {\bibfnamefont {R.}~\bibnamefont
  {Narayan}}\ and\ \bibinfo {author} {\bibfnamefont {M.~V.}\ \bibnamefont
  {Medvedev}},\ }\href@noop {} {\bibfield  {journal} {\bibinfo  {journal}
  {ApJL}\ }\textbf {\bibinfo {volume} {562}},\ \bibinfo {pages} {L129}
  (\bibinfo {year} {2001})}\BibitemShut {NoStop}%
\bibitem [{\citenamefont {{Lazarian}}(2006)}]{Lazarian06}%
  \BibitemOpen
  \bibfield  {author} {\bibinfo {author} {\bibfnamefont {A.}~\bibnamefont
  {{Lazarian}}},\ }\href {\doibase 10.1086/505796} {\bibfield  {journal}
  {\bibinfo  {journal} {"ApJL"}\ }\textbf {\bibinfo {volume} {645}},\ \bibinfo
  {pages} {L25} (\bibinfo {year} {2006})},\ \Eprint
  {http://arxiv.org/abs/arXiv:astro-ph/0608045} {arXiv:astro-ph/0608045}
  \BibitemShut {NoStop}%
\bibitem [{\citenamefont {Lazarian}\ \emph {et~al.}(2004)\citenamefont
  {Lazarian}, \citenamefont {Vishniac},\ and\ \citenamefont
  {Cho}}]{Lazarianetal04}%
  \BibitemOpen
  \bibfield  {author} {\bibinfo {author} {\bibfnamefont {A.}~\bibnamefont
  {Lazarian}}, \bibinfo {author} {\bibfnamefont {E.~T.}\ \bibnamefont
  {Vishniac}}, \ and\ \bibinfo {author} {\bibfnamefont {J.}~\bibnamefont
  {Cho}},\ }\href@noop {} {\bibfield  {journal} {\bibinfo  {journal} {ApJ}\
  }\textbf {\bibinfo {volume} {603}},\ \bibinfo {pages} {180} (\bibinfo {year}
  {2004})}\BibitemShut {NoStop}%
\bibitem [{\citenamefont {Kupiainen}(2003)}]{Kupiainen03}%
  \BibitemOpen
  \bibfield  {author} {\bibinfo {author} {\bibfnamefont {A.}~\bibnamefont
  {Kupiainen}},\ }\href@noop {} {\bibfield  {journal} {\bibinfo  {journal}
  {Ann. Henri Poincar\'{e}}\ }\textbf {\bibinfo {volume} {4, Suppl. 2}},\
  \bibinfo {pages} {S713} (\bibinfo {year} {2003})}\BibitemShut {NoStop}%
\bibitem [{\citenamefont {Eyink}(2011)}]{Eyink11}%
  \BibitemOpen
  \bibfield  {author} {\bibinfo {author} {\bibfnamefont {G.~L.}\ \bibnamefont
  {Eyink}},\ }\href@noop {} {\bibfield  {journal} {\bibinfo  {journal} {Phys.
  Rev. E}\ }\textbf {\bibinfo {volume} {83}},\ \bibinfo {pages} {056405}
  (\bibinfo {year} {2011})}\BibitemShut {NoStop}%
\bibitem [{\citenamefont {Eyink}\ and\ \citenamefont
  {Aluie}(2006)}]{EyinkAluie06}%
  \BibitemOpen
  \bibfield  {author} {\bibinfo {author} {\bibfnamefont {G.~L.}\ \bibnamefont
  {Eyink}}\ and\ \bibinfo {author} {\bibfnamefont {H.}~\bibnamefont {Aluie}},\
  }\href@noop {} {\bibfield  {journal} {\bibinfo  {journal} {Physica D}\
  }\textbf {\bibinfo {volume} {223}},\ \bibinfo {pages} {82} (\bibinfo {year}
  {2006})}\BibitemShut {NoStop}%
\bibitem [{\citenamefont {Eyink}(2009)}]{Eyink09}%
  \BibitemOpen
  \bibfield  {author} {\bibinfo {author} {\bibfnamefont {G.~L.}\ \bibnamefont
  {Eyink}},\ }\href@noop {} {\bibfield  {journal} {\bibinfo  {journal} {J.
  Math. Phys.}\ }\textbf {\bibinfo {volume} {50}},\ \bibinfo {pages} {083102}
  (\bibinfo {year} {2009})}\BibitemShut {NoStop}%
\bibitem [{\citenamefont {{Lazarian}}(2005)}]{Lazarian05}%
  \BibitemOpen
  \bibfield  {author} {\bibinfo {author} {\bibfnamefont {A.}~\bibnamefont
  {{Lazarian}}},\ }in\ \href {\doibase 10.1063/1.2077170} {\emph {\bibinfo
  {booktitle} {Magnetic Fields in the Universe: From Laboratory and Stars to
  Primordial Structures.}}},\ \bibinfo {series} {American Institute of Physics
  Conference Series}, Vol.\ \bibinfo {volume} {784},\ \bibinfo {editor} {edited
  by\ \bibinfo {editor} {\bibnamefont {{E.~M.~de Gouveia dal Pino, G.~Lugones,
  \& A.~Lazarian}}}}\ (\bibinfo {year} {2005})\ pp.\ \bibinfo {pages}
  {42--53},\ \Eprint {http://arxiv.org/abs/arXiv:astro-ph/0505574}
  {arXiv:astro-ph/0505574} \BibitemShut {NoStop}%
\bibitem [{\citenamefont {Santos-Lima}\ \emph {et~al.}(2010)\citenamefont
  {Santos-Lima}, \citenamefont {Lazarian}, \citenamefont {de~Gouveia
  Dal~Pino},\ and\ \citenamefont {Cho}}]{SantosdeLimaetal10}%
  \BibitemOpen
  \bibfield  {author} {\bibinfo {author} {\bibfnamefont {R.}~\bibnamefont
  {Santos-Lima}}, \bibinfo {author} {\bibfnamefont {A.}~\bibnamefont
  {Lazarian}}, \bibinfo {author} {\bibfnamefont {E.~M.}\ \bibnamefont
  {de~Gouveia Dal~Pino}}, \ and\ \bibinfo {author} {\bibfnamefont
  {J.}~\bibnamefont {Cho}},\ }\href@noop {} {\bibfield  {journal} {\bibinfo
  {journal} {ApJ}\ }\textbf {\bibinfo {volume} {714}},\ \bibinfo {pages} {442}
  (\bibinfo {year} {2010})}\BibitemShut {NoStop}%
\bibitem [{\citenamefont {{Lazarian}}(2011)}]{Lazarian11}%
  \BibitemOpen
  \bibfield  {author} {\bibinfo {author} {\bibfnamefont {A.}~\bibnamefont
  {{Lazarian}}},\ }\href@noop {} {\bibfield  {journal} {\bibinfo  {journal}
  {ArXiv e-prints}\ } (\bibinfo {year} {2011})},\ \Eprint
  {http://arxiv.org/abs/1108.2280} {arXiv:1108.2280 [astro-ph.GA]} \BibitemShut
  {NoStop}%
\bibitem [{\citenamefont {Petschek}(1964)}]{petschek64}%
  \BibitemOpen
  \bibfield  {author} {\bibinfo {author} {\bibfnamefont {H.}~\bibnamefont
  {Petschek}},\ }\href@noop {} {\bibfield  {journal} {\bibinfo  {journal}
  {AAS-NASA Symposium (NASA SP-50)}\ }\textbf {\bibinfo {volume} {(NASA
  SP-50)}},\ \bibinfo {pages} {425} (\bibinfo {year} {1964})}\BibitemShut
  {NoStop}%
\bibitem [{\citenamefont {Ciaravella}\ and\ \citenamefont
  {Raymond}(2008)}]{CiaravellaRaymond08}%
  \BibitemOpen
  \bibfield  {author} {\bibinfo {author} {\bibfnamefont {A.}~\bibnamefont
  {Ciaravella}}\ and\ \bibinfo {author} {\bibfnamefont {J.~C.}\ \bibnamefont
  {Raymond}},\ }\href@noop {} {\bibfield  {journal} {\bibinfo  {journal} {ApJ}\
  }\textbf {\bibinfo {volume} {686}},\ \bibinfo {pages} {1372} (\bibinfo {year}
  {2008})}\BibitemShut {NoStop}%
\bibitem [{\citenamefont {Daughton}\ \emph {et~al.}(2008)\citenamefont
  {Daughton}, \citenamefont {Roytershteyn}, \citenamefont {Albright},
  \citenamefont {Bowers}, \citenamefont {Yin},\ and\ \citenamefont
  {Karimabadi}}]{Daughtonetal08}%
  \BibitemOpen
  \bibfield  {author} {\bibinfo {author} {\bibfnamefont {W.}~\bibnamefont
  {Daughton}}, \bibinfo {author} {\bibfnamefont {V.}~\bibnamefont
  {Roytershteyn}}, \bibinfo {author} {\bibfnamefont {B.~J.}\ \bibnamefont
  {Albright}}, \bibinfo {author} {\bibfnamefont {K.}~\bibnamefont {Bowers}},
  \bibinfo {author} {\bibfnamefont {L.}~\bibnamefont {Yin}}, \ and\ \bibinfo
  {author} {\bibfnamefont {H.}~\bibnamefont {Karimabadi}},\ }\href@noop {} {}
  (\bibinfo {year} {2008}),\ \bibinfo {note} {aGU Fall Meeting
  Abstracts}\BibitemShut {NoStop}%
\bibitem [{\citenamefont {Loureiro}\ \emph {et~al.}(2009)\citenamefont
  {Loureiro}, \citenamefont {Uzdensky}, \citenamefont {Schekochihin},
  \citenamefont {Cowley},\ and\ \citenamefont {Yousef}}]{Loureiroetal09}%
  \BibitemOpen
  \bibfield  {author} {\bibinfo {author} {\bibfnamefont {N.~F.}\ \bibnamefont
  {Loureiro}}, \bibinfo {author} {\bibfnamefont {D.~A.}\ \bibnamefont
  {Uzdensky}}, \bibinfo {author} {\bibfnamefont {A.~A.}\ \bibnamefont
  {Schekochihin}}, \bibinfo {author} {\bibfnamefont {S.~C.}\ \bibnamefont
  {Cowley}}, \ and\ \bibinfo {author} {\bibfnamefont {T.~A.}\ \bibnamefont
  {Yousef}},\ }\href@noop {} {\bibfield  {journal} {\bibinfo  {journal}
  {MNRAS}\ }\textbf {\bibinfo {volume} {399}},\ \bibinfo {pages} {L146}
  (\bibinfo {year} {2009})}\BibitemShut {NoStop}%
\bibitem [{\citenamefont {Bhattacharjee}\ \emph {et~al.}(2009)\citenamefont
  {Bhattacharjee}, \citenamefont {Huang}, \citenamefont {Yang},\ and\
  \citenamefont {Rogers}}]{Bhattacharjeeetal09}%
  \BibitemOpen
  \bibfield  {author} {\bibinfo {author} {\bibfnamefont {A.}~\bibnamefont
  {Bhattacharjee}}, \bibinfo {author} {\bibfnamefont {Y.-M.}\ \bibnamefont
  {Huang}}, \bibinfo {author} {\bibfnamefont {H.}~\bibnamefont {Yang}}, \ and\
  \bibinfo {author} {\bibfnamefont {B.}~\bibnamefont {Rogers}},\ }\href@noop {}
  {\bibfield  {journal} {\bibinfo  {journal} {Phys. Plasmas}\ }\textbf
  {\bibinfo {volume} {16}},\ \bibinfo {pages} {112102} (\bibinfo {year}
  {2009})}\BibitemShut {NoStop}%
\bibitem [{Note4()}]{Note4}%
  \BibitemOpen
  \bibinfo {note} {The idea of appealing to the tearing mode as a means of
  enhancing the reconnection speed can be traced back to Refs.~\protect
  \rev@citealp {Strauss88 ShibataTanuma01}.}\BibitemShut {Stop}%
\bibitem [{\citenamefont {Diamond}\ \emph {et~al.}(1984)\citenamefont
  {Diamond}, \citenamefont {Hazeltine}, \citenamefont {An}, \citenamefont
  {Carreras},\ and\ \citenamefont {Hicks}}]{Diamondetal84}%
  \BibitemOpen
  \bibfield  {author} {\bibinfo {author} {\bibfnamefont {P.~H.}\ \bibnamefont
  {Diamond}}, \bibinfo {author} {\bibfnamefont {R.~D.}\ \bibnamefont
  {Hazeltine}}, \bibinfo {author} {\bibfnamefont {Z.~G.}\ \bibnamefont {An}},
  \bibinfo {author} {\bibfnamefont {B.~A.}\ \bibnamefont {Carreras}}, \ and\
  \bibinfo {author} {\bibfnamefont {H.~R.}\ \bibnamefont {Hicks}},\ }\href@noop
  {} {\bibfield  {journal} {\bibinfo  {journal} {Physics of Fluids}\ }\textbf
  {\bibinfo {volume} {27}},\ \bibinfo {pages} {1449} (\bibinfo {year}
  {1984})}\BibitemShut {NoStop}%
\bibitem [{\citenamefont {Sych}\ \emph {et~al.}(2009)\citenamefont {Sych},
  \citenamefont {Nakariakov}, \citenamefont {Karlicky},\ and\ \citenamefont
  {Anfinogentov}}]{Sychetal09}%
  \BibitemOpen
  \bibfield  {author} {\bibinfo {author} {\bibfnamefont {R.}~\bibnamefont
  {Sych}}, \bibinfo {author} {\bibfnamefont {V.~M.}\ \bibnamefont
  {Nakariakov}}, \bibinfo {author} {\bibfnamefont {M.}~\bibnamefont
  {Karlicky}}, \ and\ \bibinfo {author} {\bibfnamefont {S.}~\bibnamefont
  {Anfinogentov}},\ }\href@noop {} {\bibfield  {journal} {\bibinfo  {journal}
  {AandA}\ }\textbf {\bibinfo {volume} {505}},\ \bibinfo {pages} {791}
  (\bibinfo {year} {2009})}\BibitemShut {NoStop}%
\bibitem [{\citenamefont {Lapenta}(2008)}]{Lapenta08}%
  \BibitemOpen
  \bibfield  {author} {\bibinfo {author} {\bibfnamefont {G.}~\bibnamefont
  {Lapenta}},\ }\href@noop {} {\bibfield  {journal} {\bibinfo  {journal} {Phys.
  Rev. Lett.}\ }\textbf {\bibinfo {volume} {100}},\ \bibinfo {pages} {235001}
  (\bibinfo {year} {2008})}\BibitemShut {NoStop}%
\bibitem [{\citenamefont {Bettarini}\ and\ \citenamefont
  {Lapenta}(2010)}]{BettariniLapenta09}%
  \BibitemOpen
  \bibfield  {author} {\bibinfo {author} {\bibfnamefont {L.}~\bibnamefont
  {Bettarini}}\ and\ \bibinfo {author} {\bibfnamefont {G.}~\bibnamefont
  {Lapenta}},\ }\href@noop {} {\bibfield  {journal} {\bibinfo  {journal} {A \&
  A}\ }\textbf {\bibinfo {volume} {518}},\ \bibinfo {pages} {A57} (\bibinfo
  {year} {2010})}\BibitemShut {NoStop}%
\bibitem [{Lap()}]{LapentaLazarian11}%
  \BibitemOpen
  \href@noop {} {}\BibitemShut {NoStop}%
\bibitem [{\citenamefont {{Fabian}}\ \emph {et~al.}(2011)\citenamefont
  {{Fabian}}, \citenamefont {{Sanders}}, \citenamefont {{Williams}},
  \citenamefont {{Lazarian}}, \citenamefont {{Ferland}},\ and\ \citenamefont
  {{Johnstone}}}]{Fabianetal11}%
  \BibitemOpen
  \bibfield  {author} {\bibinfo {author} {\bibfnamefont {A.~C.}\ \bibnamefont
  {{Fabian}}}, \bibinfo {author} {\bibfnamefont {J.~S.}\ \bibnamefont
  {{Sanders}}}, \bibinfo {author} {\bibfnamefont {R.~J.~R.}\ \bibnamefont
  {{Williams}}}, \bibinfo {author} {\bibfnamefont {A.}~\bibnamefont
  {{Lazarian}}}, \bibinfo {author} {\bibfnamefont {G.~J.}\ \bibnamefont
  {{Ferland}}}, \ and\ \bibinfo {author} {\bibfnamefont {R.~M.}\ \bibnamefont
  {{Johnstone}}},\ }\href {\doibase 10.1111/j.1365-2966.2011.19034.x}
  {\bibfield  {journal} {\bibinfo  {journal} {"MNRAS"}\ }\textbf {\bibinfo
  {volume} {417}},\ \bibinfo {pages} {172} (\bibinfo {year} {2011})},\ \Eprint
  {http://arxiv.org/abs/1105.1735} {arXiv:1105.1735 [astro-ph.GA]} \BibitemShut
  {NoStop}%
\bibitem [{\citenamefont {{Santos-Lima}}\ \emph {et~al.}(2011)\citenamefont
  {{Santos-Lima}}, \citenamefont {{de Gouveia Dal Pino}},\ and\ \citenamefont
  {{Lazarian}}}]{SantosLimaetal11}%
  \BibitemOpen
  \bibfield  {author} {\bibinfo {author} {\bibfnamefont {R.}~\bibnamefont
  {{Santos-Lima}}}, \bibinfo {author} {\bibfnamefont {E.~M.}\ \bibnamefont {{de
  Gouveia Dal Pino}}}, \ and\ \bibinfo {author} {\bibfnamefont
  {A.}~\bibnamefont {{Lazarian}}},\ }\href@noop {} {\bibfield  {journal}
  {\bibinfo  {journal} {ArXiv e-prints}\ } (\bibinfo {year} {2011})},\ \Eprint
  {http://arxiv.org/abs/1109.3716} {arXiv:1109.3716 [astro-ph.GA]} \BibitemShut
  {NoStop}%
\bibitem [{\citenamefont {{de Gouveia dal Pino}}\ and\ \citenamefont
  {{Lazarian}}(2005)}]{deGouveiadalPinoLazarian05}%
  \BibitemOpen
  \bibfield  {author} {\bibinfo {author} {\bibfnamefont {E.~M.}\ \bibnamefont
  {{de Gouveia dal Pino}}}\ and\ \bibinfo {author} {\bibfnamefont
  {A.}~\bibnamefont {{Lazarian}}},\ }\href {\doibase
  10.1051/0004-6361:20042590} {\bibfield  {journal} {\bibinfo  {journal}
  {"A\&A"}\ }\textbf {\bibinfo {volume} {441}},\ \bibinfo {pages} {845}
  (\bibinfo {year} {2005})}\BibitemShut {NoStop}%
\bibitem [{\citenamefont {{Drake}}\ \emph {et~al.}(2009)\citenamefont
  {{Drake}}, \citenamefont {{Swisdak}}, \citenamefont {{Phan}}, \citenamefont
  {{Cassak}}, \citenamefont {{Shay}}, \citenamefont {{Lepri}}, \citenamefont
  {{Lin}}, \citenamefont {{Quataert}},\ and\ \citenamefont
  {{Zurbuchen}}}]{Drakeetal09}%
  \BibitemOpen
  \bibfield  {author} {\bibinfo {author} {\bibfnamefont {J.~F.}\ \bibnamefont
  {{Drake}}}, \bibinfo {author} {\bibfnamefont {M.}~\bibnamefont {{Swisdak}}},
  \bibinfo {author} {\bibfnamefont {T.~D.}\ \bibnamefont {{Phan}}}, \bibinfo
  {author} {\bibfnamefont {P.~A.}\ \bibnamefont {{Cassak}}}, \bibinfo {author}
  {\bibfnamefont {M.~A.}\ \bibnamefont {{Shay}}}, \bibinfo {author}
  {\bibfnamefont {S.~T.}\ \bibnamefont {{Lepri}}}, \bibinfo {author}
  {\bibfnamefont {R.~P.}\ \bibnamefont {{Lin}}}, \bibinfo {author}
  {\bibfnamefont {E.}~\bibnamefont {{Quataert}}}, \ and\ \bibinfo {author}
  {\bibfnamefont {T.~H.}\ \bibnamefont {{Zurbuchen}}},\ }\href {\doibase
  10.1029/2008JA013701} {\bibfield  {journal} {\bibinfo  {journal} {Journal of
  Geophysical Research (Space Physics)}\ }\textbf {\bibinfo {volume} {114}},\
  \bibinfo {pages} {A05111} (\bibinfo {year} {2009})}\BibitemShut {NoStop}%
\bibitem [{\citenamefont {{Lazarian}}\ and\ \citenamefont
  {{Opher}}(2009)}]{LazarianOpher09}%
  \BibitemOpen
  \bibfield  {author} {\bibinfo {author} {\bibfnamefont {A.}~\bibnamefont
  {{Lazarian}}}\ and\ \bibinfo {author} {\bibfnamefont {M.}~\bibnamefont
  {{Opher}}},\ }\href {\doibase 10.1088/0004-637X/703/1/8} {\bibfield
  {journal} {\bibinfo  {journal} {"ApJ"}\ }\textbf {\bibinfo {volume} {703}},\
  \bibinfo {pages} {8} (\bibinfo {year} {2009})},\ \Eprint
  {http://arxiv.org/abs/0905.1120} {arXiv:0905.1120 [astro-ph.EP]} \BibitemShut
  {NoStop}%
\bibitem [{\citenamefont {{Drake}}\ \emph {et~al.}(2010)\citenamefont
  {{Drake}}, \citenamefont {{Opher}}, \citenamefont {{Swisdak}},\ and\
  \citenamefont {{Chamoun}}}]{Drakeetal10}%
  \BibitemOpen
  \bibfield  {author} {\bibinfo {author} {\bibfnamefont {J.~F.}\ \bibnamefont
  {{Drake}}}, \bibinfo {author} {\bibfnamefont {M.}~\bibnamefont {{Opher}}},
  \bibinfo {author} {\bibfnamefont {M.}~\bibnamefont {{Swisdak}}}, \ and\
  \bibinfo {author} {\bibfnamefont {J.~N.}\ \bibnamefont {{Chamoun}}},\ }\href
  {\doibase 10.1088/0004-637X/709/2/963} {\bibfield  {journal} {\bibinfo
  {journal} {"ApJL"}\ }\textbf {\bibinfo {volume} {709}},\ \bibinfo {pages}
  {963} (\bibinfo {year} {2010})},\ \Eprint {http://arxiv.org/abs/0911.3098}
  {arXiv:0911.3098 [astro-ph.SR]} \BibitemShut {NoStop}%
\bibitem [{\citenamefont {{Lazarian}}\ and\ \citenamefont
  {{Desiati}}(2010)}]{LazarianDesiati10}%
  \BibitemOpen
  \bibfield  {author} {\bibinfo {author} {\bibfnamefont {A.}~\bibnamefont
  {{Lazarian}}}\ and\ \bibinfo {author} {\bibfnamefont {P.}~\bibnamefont
  {{Desiati}}},\ }\href {\doibase 10.1088/0004-637X/722/1/188} {\bibfield
  {journal} {\bibinfo  {journal} {"ApJ"}\ }\textbf {\bibinfo {volume} {722}},\
  \bibinfo {pages} {188} (\bibinfo {year} {2010})},\ \Eprint
  {http://arxiv.org/abs/1008.1981} {arXiv:1008.1981 [astro-ph.CO]} \BibitemShut
  {NoStop}%
\bibitem [{\citenamefont {{Kowal}}\ \emph {et~al.}(2011)\citenamefont
  {{Kowal}}, \citenamefont {{de Gouveia Dal Pino}},\ and\ \citenamefont
  {{Lazarian}}}]{Kowaletal11}%
  \BibitemOpen
  \bibfield  {author} {\bibinfo {author} {\bibfnamefont {G.}~\bibnamefont
  {{Kowal}}}, \bibinfo {author} {\bibfnamefont {E.~M.}\ \bibnamefont {{de
  Gouveia Dal Pino}}}, \ and\ \bibinfo {author} {\bibfnamefont
  {A.}~\bibnamefont {{Lazarian}}},\ }\href {\doibase
  10.1088/0004-637X/735/2/102} {\bibfield  {journal} {\bibinfo  {journal}
  {"ApJ"}\ }\textbf {\bibinfo {volume} {735}},\ \bibinfo {pages} {102}
  (\bibinfo {year} {2011})},\ \Eprint {http://arxiv.org/abs/1103.2984}
  {arXiv:1103.2984 [astro-ph.HE]} \BibitemShut {NoStop}%
\end{thebibliography}%

\end{document}